\documentclass[10pt,times,mathptm,conference]{IEEEtran}

\usepackage{graphicx}
\usepackage{float}
\usepackage[T1]{fontenc}
\usepackage{ae,aecompl}

\usepackage{url}

\begin{document}

\title{Non-volatile Complementary Resistive Switch-based Content Addressable Memory}

\author{\authorblockN{Omid Kavehei$^{1,~3}$, Said Al-Sarawi$^1$, Sharath Sriram$^2$, Madhu Bhaskaran$^2$, Kyoung-Rok Cho$^3$, \\ Kamran~Eshraghian$^3$, and Derek Abbott$^1$}
\authorblockA{$^1~$School of Electrical \& Electronic Engineering, The University of Adelaide, SA 5005, Australia\\
$^2~$Functional Materials and Microsystems Research Group, RMIT University, Melbourne, VIC 3001, Australia\\
$^3~$Department of Electrical and Computer Engineering, Chungbuk National University, 361-763 South Korea\\
Email: \{omid, alsarawi, dabbott\}@eleceng.adelaide.edu.au \\ ~~~~~~~\{sharath.sriram, madhu.bhaskaran\}@rmit.edu.au \\  ~~~~~~~~~~~~~~~~~~~k.eshraghian@innovationlabs.com.au and krcho@cbnu.ac.kr}}

\maketitle

\begin{abstract}
This paper presents a novel resistive-only Binary and Ternary Content Addressable Memory (B/TCAM) cell that consists of two Complementary Resistive Switches (CRSs). The operation of such a cell relies on a logic$\rightarrow$ON state transition that enables this novel CRS application. 

\end{abstract}


%
\IEEEpeerreviewmaketitle

\section{Introduction}\label{intro} 

\begin{figure}[htbp] \centering
  \includegraphics[width=0.3\textwidth]{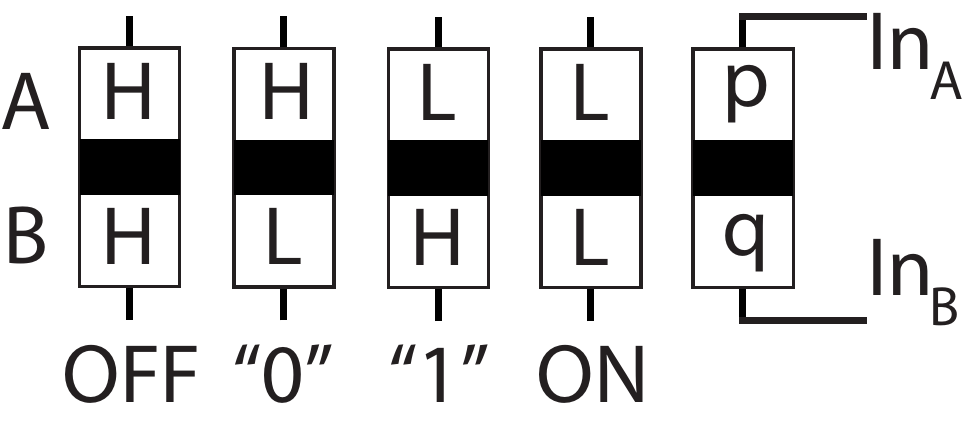}
  \caption{CRS device structure and logical definition of each combination.} \label{fig:crs_states}
\end{figure}

Linn~\emph{et al.}~\cite{crs:linn:2010} introduced a new paradigm by exploiting two serially connected non-volatile resistive memory devices (memristors~\cite{kavehei2011fabrication}) with opposite polarities. The structure uses a High Resistance State (HRS) and a Low Resistance State (LRS) to encode either logic ``0" or logic ``1". Consequently, the overall resistance of such device is always around HRS, resulting in significant reduction in the parasitic current paths through neighbouring devices, known as the sneak-paths currents. This device is referred to as a Complementary Resistive Switch (CRS). Fig.~\ref{fig:crs_states} summaries the CRS functionality. If $p$ and $q$ indicate resistances of the memristors A and B, respectively, four different states can be observed. For example, $p/q\leftarrow$L/H indicates LRS for $p$ (memristor A) and HRS for $q$ (memristor B), which is selected as a logic ``1''. Note that the H/H state only occurs once in a fresh device. Transitions between the other three states depend on the potential difference across the device terminals. These transitions can be identified by two SET and two RESET thresholds. Absolute values of these thresholds are approximately similar~\cite{crs:linn:2010,analytical:kavehei:2011}. 

Two transitions are possible: logic$\rightarrow$logic and logic$\rightarrow$ON. The first transition indicates a change from L/H (``1'') to H/L (``0'') or vice versa. This transition needs a high applied voltage $>V_{\rm th, RESET}$ that results a current spike. While a logic$\rightarrow$ON transition requires a voltage that lies between the SET and the RESET thresholds. These transitions conditionally occur depending on the device initial state and the polarity of applied voltage. This condition can be defined as a new fundamental logic operation, known as the `material implication', IMP,~\cite{borghetti:memristive:2010,kavehei:2011:memristor,kvatinsky2011memristor}. This logic operation results in change in $q$ depending on the state of $p$ that is shown as $p$ IMP $q$, `$p$ implies $q$' or `if $p$ then $q$'. Rosezin~\emph{et al.}~\cite{crs:rosezin:2011} introduced crossbar logic via material implication that we call logic$\rightarrow$logic implication. While here we introduce and use a logic$\rightarrow$ON implication for CRS devices. 

A logic$\rightarrow$ON transition occur if $V_{\rm th, SET}<V($In$_{\rm A})$-$V($In$_{\rm B})<V_{\rm th, RESET}$, and $p\leftarrow$L \& $q\leftarrow$H (logic ``1''). Under these conditions state of $q$ changes to L, which shows that $p$ IMP $q$. In this fashion, the device operation can be considered as a Mealy finite state machine whose output value(s) are determined both by its current state and by inputs. Note that, here $p$ and $q$ are resistance states, whereas in~\cite{crs:rosezin:2011} they are input signals. In this situation, $p\leftarrow$L \& $q\leftarrow$L, a significantly large current can be driven through the device compare to a logic state. The logic$\rightarrow$ON implication is the basic operation to design the CRS-based CAM cell.    

\begin{figure*}[htbp]
\centering
  \includegraphics[width=0.8\textwidth]{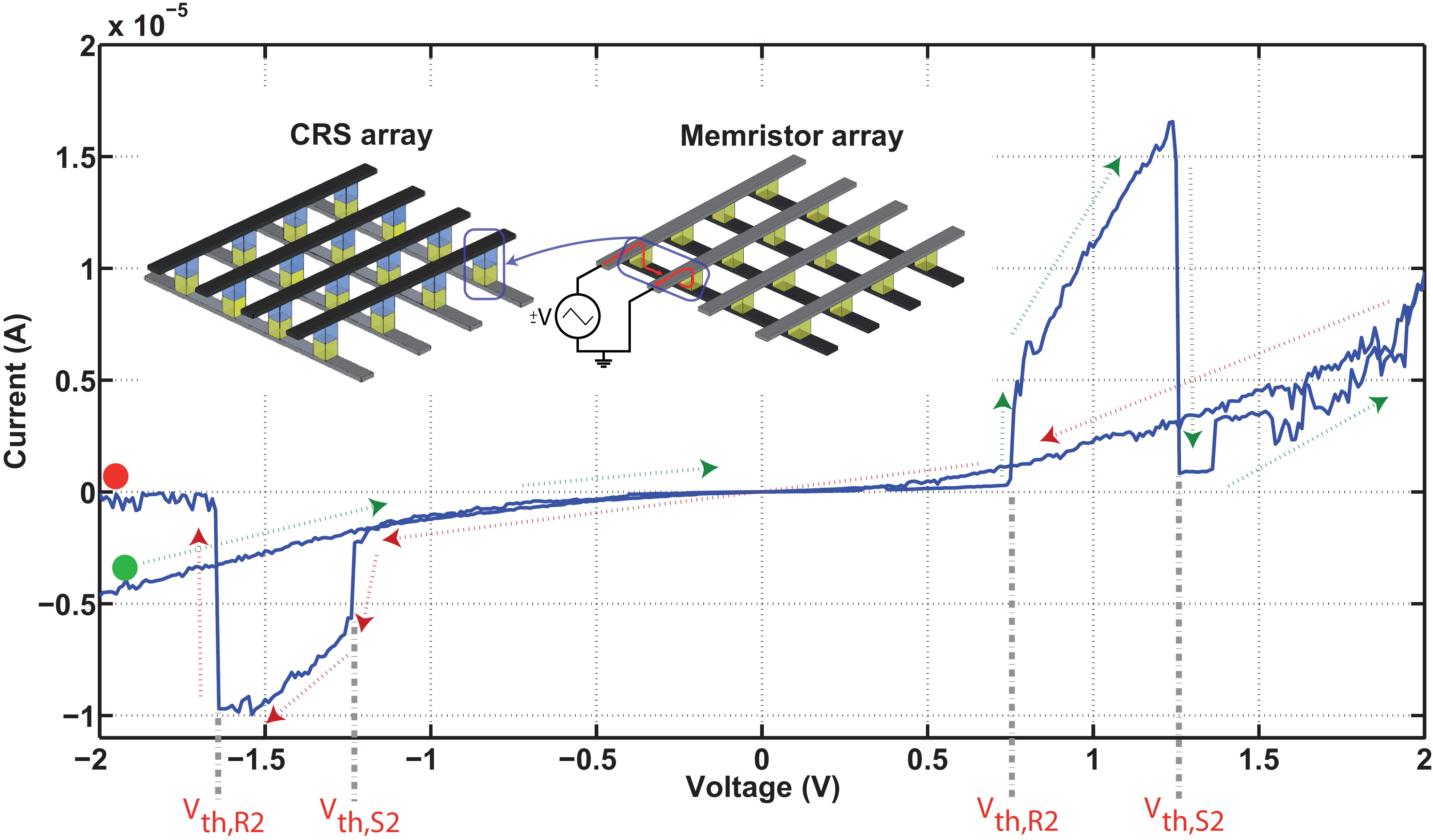}
  \caption{CRS device measurement using two memristor devices connected in series with no control applied to the middle electrode. The green dot demonstrates starting point and the red dot shows final point of the experiment. Inset illustrates the memristor array and the implementation of a CRS device characteristics by applying $-2$~V to $+2$~V voltage range. Fabrication steps and further details are available in \cite{kavehei2011fabrication}. The red path in the inset shows the current.}\label{fig:crs_meas}
\end{figure*}

Fig.~\ref{fig:crs_meas} illustrates a behaviour similar to a CRS device using two formed memristors. This measurement highlights that two bipolar memristors connected in series with opposite polarities behave as a CRS device. A fully integrated CRS $I$-$V$ characteristics will be more symmetrical. The asymmetry of the $I$-$V$ curve is mainly due to: (i) using two different memristors instead of a single CRS device, (ii) different contact sizes of the two memristors, (iii) and possibly fluctuations that affect memristive behaviour during the initial irreversible electroforming process for the memristors. The structure is like a memistor (note the missing ``r'') device~\cite{memistor:widrow:1960}. A possible fabrication technique of a three-terminal resistive switch (memistor) is introduced in~\cite{memistor:xia:2011}. 
 
The functionality of a CRS device is summarised in Table~\ref{tab:crs_states-trans}, where $R'$ shows the next resistance state, $R$ illustrates the initial resistance state, and output is a current pulse or spike. After applying a positive or negative bias, depending on the polarity of memristors, the device switches to either the ``0" or ``1" state. In Fig.~\ref{fig:crs_meas}, the dashed Gray lines are threshold voltages for SET, $V_{\rm th,S1}$ and $V_{\rm th,S2}$, and RESET, $V_{\rm th,R1}$ and $V_{\rm th,R2}$. In an ideal case,  $V_{\rm th,SET}=V_{\rm th,S1}=|V_{\rm th,S2}|$ and $V_{\rm th,RESET}=V_{\rm th,R1}=|V_{\rm th,R2}|$. A successful READ operation occurs if $V_{\rm th,SET}<V_{\rm READ}<V_{\rm th,RESET}$. For a successful WRITE, $V_{\rm th,RESET}<V_{\rm WRITE}$. Consequently, every voltage below $V_{\rm th,SET}$ should not contribute any change in the device state.  

\begin{table}[htpb] \centering
    \caption{State transitions in CRS}
    \label{tab:crs_states-trans}

    \begin{small}
    \begin{tabular}{|c|c|c|c|}
    \hline \hline
		{$R$} & {$\Delta V$} & {$R'$} & {Output} \\ \hline\hline
		{High (``1'')} & {$V_{\rm th,S1}<\Delta V<V_{\rm th,R1}$} & {Low ~(ON)} & {pulse} \\ \hline
		{High (``1'')} & {$V_{\rm th,R1}<\Delta V$} & {High (``0'')} & {spike} \\ \hline
		{High (``0'')} & {$V_{\rm th,R2}<\Delta V<V_{\rm th,S2}$} & {Low ~(ON)} & {pulse} \\ \hline
		{High (``0'')} & {$\Delta V<V_{\rm th,R2}$} & {High (``1'')} & {spike} \\ \hline
		{Low ~(ON)} & {$V_{\rm th,R1}<\Delta V$} & {High (``0'')} & {--} \\ \hline
		{Low ~(ON)} & {$\Delta V<V_{\rm th,R2}$} & {High (``1'')} & {--} \\ \hline
		\end{tabular}
    \end{small} 
\end{table}

There are numerous applications for CRS-based B/TCAM, including pattern matching applications in real-time network intrusion detection, network packet routing, DNA sequencing~\cite{alibart2011hybrid}. Inclusion of CRS devices within a three dimensional hybrid CMOS/nanodevice architecture enables improvements in throughput, density, and power performance relative to state-of-the-art designs \cite{alibart2011hybrid,eshraghian2010memristor}.

\section{Proposed CRS-based CAM Cell}\label{proposed} 
\begin{figure}[htbp] \centering
  \includegraphics[width=0.48\textwidth]{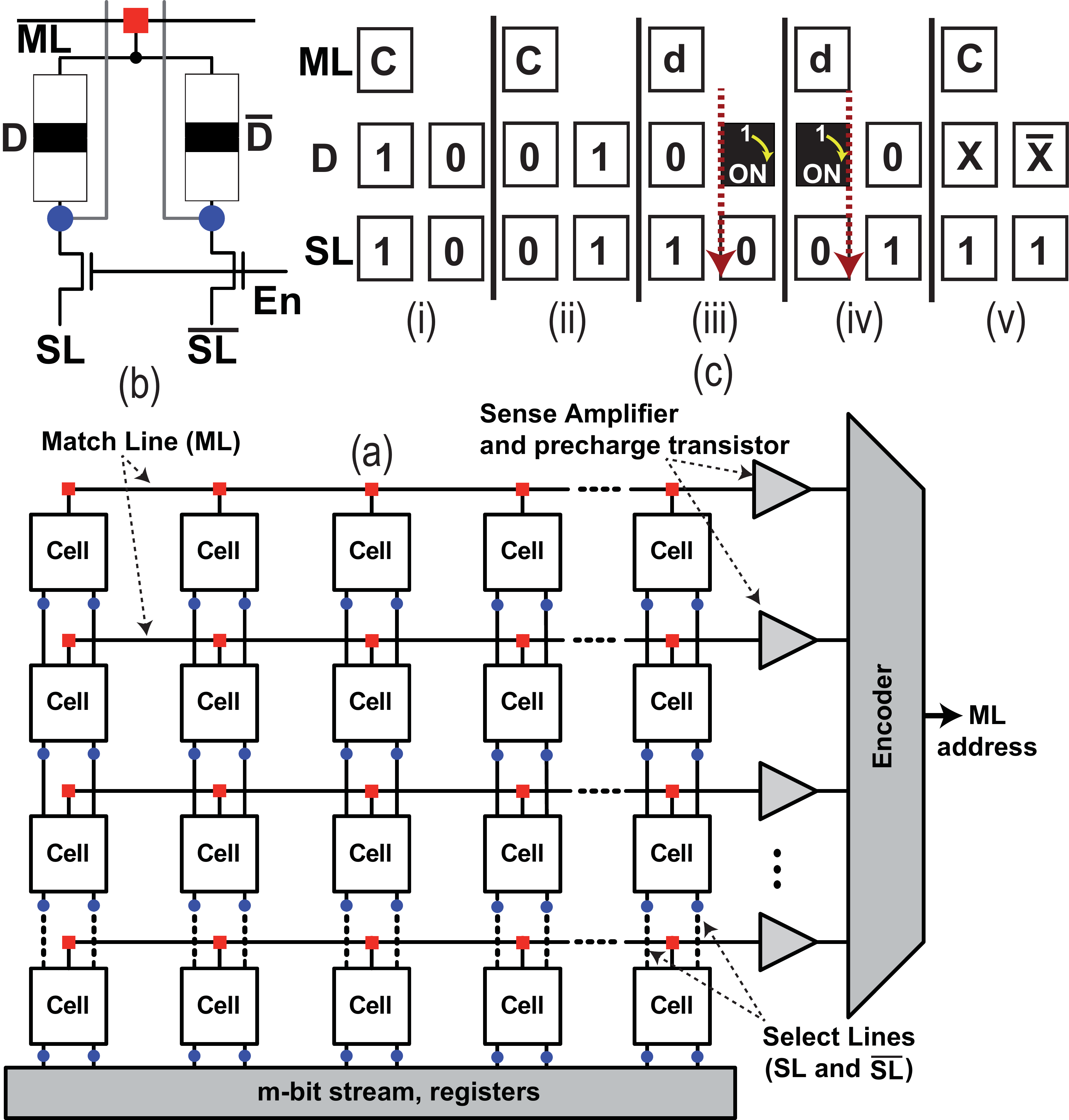}
  \caption{The CAM system, (a), and proposed cell (b). Design is compatible with CMOL architecture. The blue and red via contacts between the CMOS FPGA type array and CRS-based array highlights this compatibility. The logical operation of the B/TCAM cell is shown in (c). ${\rm ML}=$C and d demonstrate charged and discharged ${\rm ML}$, respectively. Further information regarding the density and the connections (via) between the nano and CMOS domain can be found in~\cite{alibart2011hybrid}.} \label{fig:arch}
\end{figure}

The proposed structure uses the in-situ computing  of memristor and CRS devices, as described in~\cite{kavehei:2011:memristor}. Here, however, it is compatible with the four-dimensional CMOL (CMOS MOLecular scale devices) architecture that is described in~\cite{defect:strukov:2007} and as its unique feature, it contains CRS devices that is stacked on top of the CMOS layer(s). 
Fig.~\ref{fig:arch} summarises the idea. The grey and white modules in Fig.~\ref{fig:arch}~(a) represent the CMOS domain and the nano domain implementations, respectively. The red (rectangular) and blue (circle) dots correspond to via connections from the CMOS domain to the nano domain. The implementation is compatible with the CMOL architecture using a Field Programmable Gate Array (FPGA) type structure~\cite{defect:strukov:2007}. The trade-off analysis between other methods in implementing the interface between the CMOS and the nano domain is beyond the scope of this letter. 
Each cell represents a CAM cell that is entirely implemented in the nano domain. 
Fig.~\ref{fig:arch}~(b) illustrates the CAM cell structure that works for binary CAM and a more flexible type, ternary CAM, applications. For simplicity, the dot colours are chosen to be identical with the via colours in~\cite{defect:strukov:2007}. It has to be stated that connecting the middle electrode (the black line in the CRS symbol) is not allowed because it, firstly, creates new sneak-path currents and, secondly, affects the expected functionality of CRS devices. 

Fig.~\ref{fig:arch}~(c) demonstrates the cell functionality. Possible combinations are shown in Fig.~\ref{fig:arch}~(c)-(i)~to~-(v). If the stored data (${\rm D}$) in the complementary cell (consisting of ${\rm D}$ and $\bar{\rm D}$) is ``1'' and it is matched with the complementary select lines (${\rm SL}$ and $\bar{\rm SL}$)  during the evaluation phase (active ${\rm En}$), no path is between the pre-charged ${\rm ML}$ and ${\rm SL}$s. Likewise, if ${\rm D}={\rm SL}=$``0'', ${\rm ML}$ remains charged (for ${\rm D}$ and ${\rm SL}$ vectors all the elements must be matched).  The only possibility to discharge ${\rm ML}$ is the in situation that has been defined as logic$\rightarrow$ON implication. This situation can only happen either when ${\rm D}=$``1'' and ${\rm SL}=$``0'' or ${\rm D}=$``0'' and ${\rm SL}=$``1''. This path is shown with the dashed red arrows and the corresponding ON state device is also highlighted. Either of the combinations indicates a mismatched situation a voltage drop occurs on the corresponding ${\rm ML}$. Details of the crossbar design and applied voltages are described in~\cite{analytical:kavehei:2011}. 

The proposed cell can handle TCAM operation. If a protocol for stored ``don't care'' (${\rm X}$) bits can be approved as applying a high voltage (``1'') to the both select lines when bypassing the stored bit is desirable, the logic$\rightarrow$ON conditions will never be satisfied and the device keeps its stored data. 
 
\begin{figure}[htbp] \centering
  \includegraphics[width=0.48\textwidth]{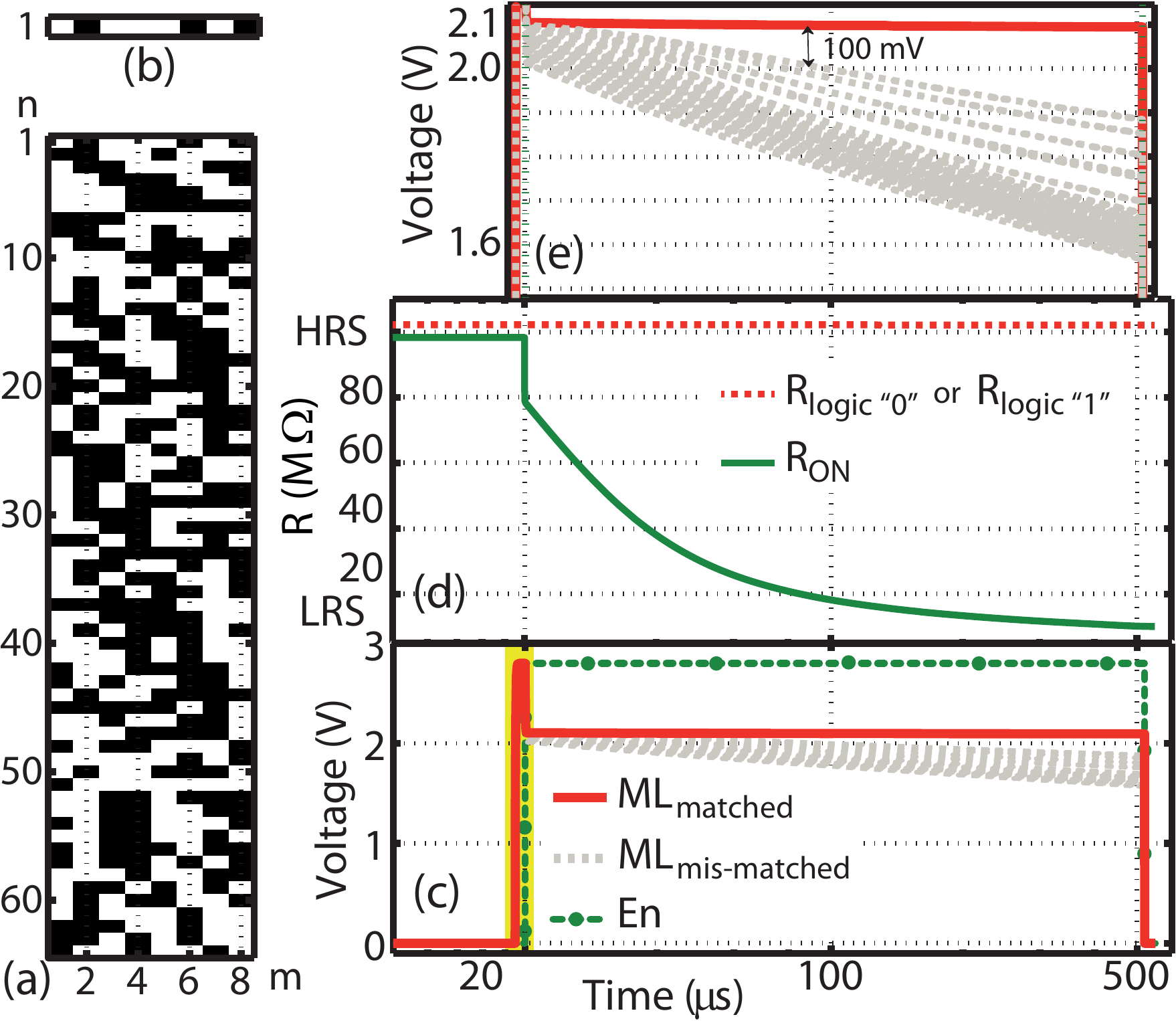}
  \caption{Simulation results using Cadence Spectre. The stored and input patterns are shown in (a) and (b), respectively. (c) demonstrates  matched and mismatched ${\rm ML}$s and the evaluation, ${\rm En}$, signal. The yellow region highlights a pre-charge step. (e) demonstrates a more clear picture of (c). Approximately, $80~\mu$s for the worst-case ${\rm ML}$ to reaches the minimum detectable $\Delta V$. (d) illustrates two resistor samples (logic -either ``0'' or ``1''- and ON states).} \label{fig:sims}
\end{figure}

A $64\times 8$ CAM is simulated and its function is confirmed. In this implementation $16$ select lines and $64$ match lines are required. Fig.~\ref{fig:sims} illustrates simulation inputs and results. The stored memory pattern and the search stream are randomly generated and shown in Fig.~\ref{fig:sims}~(a) and~(b). For simplicity, in the presentation, we intentionally addressed only one (the $1^{\rm st}$) match line, therefore, the expectation is to see ${\rm ML}_{1}$ remains unchanged while all other match lines are dropped. The voltage drop must be detectable by the sense amplifier (SA) chain~\cite{analytical:kavehei:2011}. Nano-wire parasitic resistors are also taken into account.  

Figs.~\ref{fig:sims}~(c) and~(e) illustrate outputs of matched (solid line) and mismatched (dashed lines) ${\rm ML}$s and the evaluation, ${\rm En}$, signal (dashed line with circle symbols). The only match line that remains unchanged after activating the enable is the red one. The voltage drop right after activating ${\rm En}$ occurs due to the fact that each ${\rm ML}$ acts as the middle node of a voltage divider with an effective pull-up and pull-down resistors. Therefore, another advantage of CRS-based CAM is that the pull-down resistors are initially $R_{\rm HRS}$, whereas a memristor-based array is highly pattern dependent.   
Fig.~\ref{fig:sims}~(d) demonstrates resistances of two CRS devices, with stored logic ``0'' or ``1'' (dashed line) and ON states (solid line). In the mismatched cells a logic$\rightarrow$ON transition is observed. The ON resistance of a CRS device is equivalent to $2R_{\rm LRS}$. The outputs then feed into an array of SAs through the red via connections. The design of these SAs requires a detail analysis of the array output for finding optimum values for crossbar memory parameters~\cite{analytical:kavehei:2011}.   

Increasing access time and decreasing energy dissipation of  of B/TCAMs are two trends that have been aggressively pursued. Although, from the fact that comparing today's mature or advanced technologies with the emerging technologies is not quite fair, but a switching speed and energy analysis of fabricated devices in~\cite{analytical:kavehei:2011,borghetti:memristive:2010} illustrates that applying higher voltage pulses exponentially increases the switching speed and it reduces overall energy dissipation. It is also observed that more than $80\%$ of the total power is consumed by the nanowires and the device itself consumes $10$-$100$ pJ dynamic energy ($30$~ns switching time)~\cite{borghetti:memristive:2010}, which is not an outstanding result compared to low-power B/TCAMs. Resistive memories and in particular, CRS devices, demonstrate relatively robust operation, non-volatile memory, high scalability, and promising experimental results for reducing switching time, while maintaining a relatively switching energy, which make them a serious alternative to the conventional CMOS technology~\cite{crs:linn:2010,analytical:kavehei:2011,crs:rosezin:2011}. In addition, as the CRS technology matures and the advanced transistor technologies continue to face more challenges, the combination of these two technologies will result in significantly more efficient and denser designs~\cite{defect:strukov:2007}.

\section{Conclusions and Future Directions}\label{conclusion} 
This paper introduced a CRS-based B/TCAM cell. The proposed cell has been mapped on a CMOL type architecture by introducing via contacts between the CMOS and the nano domains. Relatively robust, low-power, and scalable features of the CRS cell along with the CMOL's architectural flexibility introduce several advantages for the proposed cell. The cell's operation relies on the logic$\rightarrow$ON state transition that also introduces a novel application of this CRS operation. The CRS-based B/TCAM and the application of the logic$\rightarrow$ON transition have potential impact for stimulating pioneering work in CAM applications and CRS-based computing. The projected aim is to design and fabricate a CRS array utilising a non-destructive read-out technique, similar to \cite{tappertzhofen:2011:capacity}, to enhance overall performance of the memory array and significantly reduce the total number of refreshing cycles. This reduction is an important factor since the endurance requirement in for CRS devices will be relaxed with less refreshing cycles and it may be further relaxed with the increase of memory density.

\bibliographystyle{ieeetran}

\end{document}